\documentclass[preprint,nofootinbib]{revtex4}%
\usepackage{amssymb}
\usepackage{amsfonts}
\usepackage{amsmath}
\usepackage{amsmath}
\usepackage{graphicx}
\usepackage{hyperref}
\hypersetup{colorlinks,linkcolor={blue},citecolor={blue},urlcolor={black}} 

\usepackage[usenames]{color}%
\providecommand{\U}[1]{\protect\rule{.1in}{.1in}}
\providecommand{\U}[1]{\protect\rule{.1in}{.1in}}
\definecolor{blue}{rgb}{0,0,1}

\definecolor{red}{rgb}{1,0,0}

\begin{document}
\title{Exact scalar (quasi-)normal modes of black holes and solitons in gauged SUGRA}
\author{Monserrat Aguayo, Ankai Hern\'{a}ndez, Jos\'{e} Mena, Julio Oliva, Marcelo Oyarzo}
\affiliation{Departamento de F\'isica, Universidad de Concepci\'on, Casilla, 160-C,
Concepci\'on, Chile.}
\email{maguayo2018@udec.cl; ankaihercas@gmail.com; jomena@udec.cl; julioolivazapata@gmail.com; moyarzo2016@udec.cl}

\begin{abstract}
In this paper we identify a new family of black holes and solitons that lead
to the exact integration of scalar probes, even in the presence of a
non-minimal coupling with the Ricci scalar which has a non-trivial profile. The backgrounds are planar and
spherical black holes as well as solitons of $SU\left(  2\right)  \times
SU\left(  2\right)  $ $\mathcal{N}=4$ gauged supergravity in four dimensions.
On these geometries, we compute the spectrum of (quasi-)normal modes for the
non-minimally coupled scalar field. We find that the equation for the radial
dependence can be integrated in terms of hypergeometric functions leading to
an exact expression for the frequencies. The solutions do not asymptote to a
constant curvature spacetime, nevertheless the asymptotic region acquires an
extra conformal Killing vector. For the black hole, the scalar probe is purely
ingoing at the horizon, and requiring that the solutions lead to an extremum
of the action principle we impose a Dirichlet boundary condition at infinity.
Surprisingly, the quasinormal modes do not depend on the radius of the black
hole, therefore this family of geometries can be interpreted as isospectral in
what regards to the wave operator non-minimally coupled to the Ricci scalar.
We find both purely damped modes, as well as exponentially growing unstable
modes depending on the values of the non-minimal coupling parameter. For the
solitons we show that the same integrability property is achieved separately
in a non-supersymmetric solutions as well as for the supersymmetric one.
Imposing regularity at the origin and a well defined extremum for the action principle
we obtain the spectra that can also lead to purely oscillatory modes as well
as to unstable scalar probes, depending on the values of the non-minimal coupling.

\end{abstract}
\maketitle

\section{Introduction}

Quasinormal modes play a very important role both in astrophysical as well as
in theoretical contexts. In the former, they dominate the ringdown dynamics
of the final black hole obtained from the fusion of compact objects, and a direct measurement of the mode with the lowest damping helps
obtaining the mass and angular momentum of the final object
\cite{LIGOScientific:2016aoc}. In the latter, black hole quasinormal modes, within the
context of holography, allow for the computation of relaxation properties of
the dual field theory living at the boundary of AdS
\cite{Horowitz:1999jd},\cite{Birmingham:2001pj}. It is well-known that even for simple black
holes, as for example for Schwarzschild-(A)dS the computation of quasinormal
modes relies on numerical techniques. These techniques are fully reliable,
notwithstanding there are particular interesting cases where the spectrum of
quasinormal frequencies can be found analytically which are useful to explore the relaxation properties of
perturbations outside a black hole in an exact manner as one modifies the
parameters that define the background geometry. A partial list of such cases
is given by \cite{Chan:1996yk}-\cite{Chernicoff:2020kmf}. In this paper, we identify a new family of
black holes and solitons that allow for the exact integration of non-minimally
coupled scalar probes, in the context of $SU\left(  2\right)  \times SU\left(
2\right)  $ $\mathcal{N}=4$ gauged supergravity in four dimensions. This
theory, also known as the Freedman-Schwarz \cite{FS} model can be obtained from 10D
supergravity compactified on $S^{3}\times S^{3}$ \cite{CVcorto}, \cite{CVlargo}. The action
principle reads%
\begin{align}
S &  =\int d^{4}x\sqrt{-g}\left[  \frac{R}{4}-\frac{1}{2}\partial_{\mu}%
\phi\partial^{\mu}\phi-\frac{1}{2}e^{4\phi}\partial_{\mu}a\partial^{\mu
}a+\frac{e_{A}^{2}+e_{B}^{2}}{8}e^{2\phi}-\frac{e^{-2\phi}}{4}\left(
A^{i\mu\nu}A_{i\mu\nu}+B^{i\mu\nu}B_{i\mu\nu}\right)  \right.
\nonumber\\
&  \left.  -\frac{\mathbf{a}}{4}\frac{\epsilon^{\mu\nu\rho\sigma}}{\sqrt{-g}%
}\left(  A_{i\mu\nu}A_{\ \rho\sigma}^{i}+B_{i\mu\nu}B_{\ \rho\sigma}%
^{i}\right)  \right]  \label{action sugra}
\end{align}
and the axion field $\mathbf{a}$ can consistently be set to zero provided the
following equation of Pontryagin densities for the $SU\left(  2\right)  $
gauge fields holds%
\begin{equation}
\epsilon^{\mu\nu\rho\sigma}\left(  A_{i\mu\nu}A_{\ \rho\sigma}^{i}+B_{i\mu\nu
}B_{\ \rho\sigma}^{i}\right)  =0\ .
\end{equation}

The field strength for the gauge fields are given by%
\begin{equation}
A_{\mu\nu}^{i}=\partial_{\mu}A_{\nu}^{i}-\partial_{\nu}A_{\mu}^{i}%
+e_{A}\epsilon_{ijk}A_{\mu}^{j}A_{\nu}^{k}\text{\qquad and\qquad}B_{\mu\nu
}^{i}=\partial_{\mu}B_{\nu}^{i}-\partial_{\nu}B_{\mu}^{i}+e_{B}\epsilon
_{ijk}B_{\mu}^{j}B_{\nu}^{k}\ .
\end{equation}

\bigskip

In this work we will focus on the computation of quasinormal modes of scalar
probes on black holes of this theory as well as on the computation of normal
frequencies of the same probe fields on the gravitational soliton recently
constructed in \cite{Canfora:2021nca}, both in the supersymmetric and non-supersymmetric
cases. We will deal with solutions with vanishing axion field, and since
the self-interaction of the dilaton does not have a local extremum, the
solutions have an asymptotic structure that has less symmetry than a maximally
symmetric spacetime, although we will see the emergence of an asymptotic conformal Killing vector.

\bigskip

\section{Scalar probes on black holes}

The two families of black holes we will be interested in this section were
constructed in \cite{Klemm}. The metric in both cases, namely spherical and planar,
reads%
\begin{equation}
ds^{2}=-\alpha r^{2}\left(  1-\frac{r_{+}^{2}}{r^{2}}\right)  dt^{2}%
+\frac{dr^{2}}{\alpha\left(  1-\frac{r_{+}^{2}}{r^{2}}\right)  }+r^{2}%
d\Sigma_{2}^{2}\ , \label{bh}%
\end{equation}
where $\Sigma_{2}$ is a two-dimensional Euclidean manifold of constant
curvature $\gamma=+1,0$.

In the spherically symmetric case, $\gamma=+1$ and $d\Sigma_{2}^{2}%
=d\theta^{2}+\sin^{2}\theta d\varphi^{2}$ is the line element of the round
two-sphere, while the constant $\alpha$, the dilaton and gauge fields read%
\begin{align}
\alpha &  =\frac{1}{2}\left(  e_{A}^{2}+e_{B}^{2}\right)  \left(  H_{A}%
^{2}+H_{B}^{2}\right)  +\frac{1}{4}\ ,\\
\phi\left(  r\right)   &  =-\ln\left(  \frac{r}{2\sqrt{H_{A}^{2}+H_{B}^{2}}%
}\right)  \ ,\\
A_{\left[  1\right]  }^{i}  &  =-H_{A}\cos\theta d\varphi\delta_{3}^{i}\ ,\\
B_{\left[  1\right]  }^{i}  &  =-H_{B}\cos\theta d\varphi\delta_{3}^{i}\ .
\end{align}

In the planar case, $\gamma=0$, $d\Sigma_{2}^{2}=dx^{2}+dy^{2}$ the gauge
fields vanish and%
\begin{align}
&  \alpha=\frac{e_{A}^{2}+e_{B}^{2}}{8}\ ,\\
&  \phi\left(  r\right)  =-\ln\left(  r\right)  \ .
\end{align}

The black holes (\ref{bh}) approach the background%

\begin{equation}
ds_{\text{back}}^{2}=-\alpha r^{2}dt^{2}+\frac{dr^{2}}{\alpha}+r^{2}%
d\Sigma_{2}^{2}\ ,\label{back}%
\end{equation}
with the following asymptotic behavior%
\begin{equation}
\delta g_{tt}=\mathcal{O}\left(  1\right)  \text{, }\delta g_{rr}%
=\mathcal{O}\left(  r^{-2}\right)  \ .
\end{equation}
Notice that the background (\ref{back}) has an extra conformal Killing vector
generated by $r\rightarrow\lambda r$. The temperature of this black hole has
the intriguing property of being independent of the $r_{+}$, namely a
constant, and it is given by%
\begin{equation}
T=\frac{\alpha}{2\pi}\ .
\end{equation}
As we show below, a similar feature occurs with the quasinormal frequencies of
the non-minimally coupled scalar on this geometry, which do not depend on
$r_{+}$, leading to isospectral geometries in what regards to such operator.
Wald's formula for the entropy yields%
\begin{equation}
S=\frac{A}{4G}=\pi r_{+}^{2}\text{Vol}\left(  \Sigma\right)  \ ,
\end{equation}
where Vol$\left(  \Sigma\right)  $ is the volume of the Euclidean manifold
$\Sigma_{2}$ and we have normalized the Einstein term in the action
(\ref{action sugra}) such that $G=(4\pi)^{-1}$. First law%
\begin{equation}
dM=TdS\ ,
\end{equation}
provides the following value for the mass of the black hole%
\begin{equation}
M=\frac{\alpha r_{+}^{2}\text{Vol}\left(  \Sigma\right)  }{2}\ .
\end{equation}

\bigskip

Here, as an avatar for the study of the stability of these black holes, we
will consider a real scalar probe, coupled to the Ricci scalar in a
non-minimal manner:%
\begin{equation}
\square\Phi-\xi R\Phi=0\ , \label{scalar}%
\end{equation}
on the background (\ref{bh}).

\bigskip

Given the local isometries of the spacetime, the scalar probe admits a mode
separation which is given by%
\begin{equation}
\Phi\left(  t,r,y^{i}\right)  =\operatorname{Re}\left(  \int d\omega\sum
_{A}e^{-i\omega t}H_{\omega,A}\left(  r\right)  Y_{A}\left(  y\right)
\right)  \ , \label{separation}%
\end{equation}
where $y^{i}$ are the coordinates on the Euclidean manifold $\Sigma_{2}$ and
$Y_{k}\left(  y\right)  $ are harmonic function on such manifold, which are
labeled by the multi-index $A$. Concretely, for the spherically symmetric case
the harmonic functions are standard spherical harmonics, namely $A=\left\{
l,m\right\}  $ and they fulfil%
\begin{equation}
\nabla_{S^{2}}^{2}Y_{l,m}=-k^{2}Y_{l,m}=-l\left(  l+1\right)  Y_{l,m}\ ,
\end{equation}
while for the planar case, the harmonic functions are trivially given by plane
waves of the form%
\begin{equation}
Y_{A}=Y_{\vec{k}}=Ce^{-i\vec{k}\cdot\vec{y}}\ ,
\end{equation}
which fulfil%
\begin{equation}
\nabla_{R^{2}}Y_{\vec{k}}=-k^{2}Y_{\vec{k}}=-\left(  k_{1}^{2}+k_{2}%
^{2}\right)  Y_{\vec{k}}\ .
\end{equation}
Hereafter, for brevity we introduce the notation $H_{\omega,A}\left(
r\right)  =H\left(  r\right)  $.

\bigskip

Notice that since the Ricci scalar of the spacetime has a non-trivial radial
profile%
\begin{equation}
R=\frac{2\gamma-6\alpha}{r^{2}}-\frac{2\alpha r_{+}^{2}}{r^{4}}\ ,
\end{equation}
the non-minimal coupling term in (\ref{scalar}) cannot be seen as an effective
mass term. In spite of this fact, we will show that the equation for the
radial profile of the scalar probe $H\left(  r\right)  $ can be solved in an
exact manner in terms of hypergeometric functions.

\bigskip

Introducing the separation (\ref{separation}) on the scalar field equation
(\ref{scalar}) as a probe field on the black hole metric (\ref{bh}), after
performing the change of variables%
\begin{equation}
r=\frac{r_{+}}{\left(  1-x\right)  ^{1/2}}\ ,
\end{equation}
which maps the region of outer communication $r\in\lbrack r_{+},+\infty
\lbrack$ to $x\in\lbrack0,1[$, leads to the following equation for the radial
profile%
\begin{equation}
\frac{d^{2}H\left(  x\right)  }{dx^{2}}+\frac{1}{x}\frac{dH\left(  x\right)
}{dx}+\left(  \frac{\omega^{2}}{4\alpha^{2}x^{2}\left(  1-x\right)  ^{2}%
}-\frac{k^{2}}{4\alpha x\left(  1-x\right)  ^{2}}-\frac{(\alpha x-4\alpha
+\gamma)}{2\alpha x\left(  1-x\right)  ^{2}}\xi\right)  H\left(  x\right)
=0\ .
\end{equation}
Remarkably, this equation admits a solution in terms of hypergeometric
functions. After imposing ingoing boundary condition at the horizon one
obtains%
\begin{equation}
H\left(  x\right)  =C_{1}x^{-\frac{i\omega}{2\alpha}}(1-x)^{\frac{\alpha
+\sqrt{\left(  1-6\xi\right)  \alpha^{2}+2\xi\alpha\gamma+\alpha k^{2}%
-\omega^{2}}}{2\alpha}}F\left(  a_{1},b_{1},c_{1},x\right)
\ ,\label{ingoingnonminimal}%
\end{equation}
with%
\begin{align}
a_{1} &  =\frac{1}{2}-\frac{i\omega}{2\alpha}-\frac{\sqrt{2\xi}}{2}%
+\frac{\sqrt{\left(  1-6\xi\right)  \alpha^{2}+2\xi\alpha\gamma+\alpha
k^{2}-\omega^{2}}}{2\alpha}\ ,\\
b_{1} &  =\frac{1}{2}-\frac{i\omega}{2\alpha}+\frac{\sqrt{2\xi}}{2}%
+\frac{\sqrt{\left(  1-6\xi\right)  \alpha^{2}+2\xi\alpha\gamma+\alpha
k^{2}-\omega^{2}}}{2\alpha}\ ,\\
c_{1} &  =1-\frac{i\omega}{\alpha}\ .
\end{align}

Using Kummer identities, the ingoing solution (\ref{ingoingnonminimal}) can be
rewritten as%
\begin{align}
H\left(  x\right)   &  =C_{1}x^{-\frac{i\omega}{2\alpha}}(1-x)^{\frac
{\alpha+\sqrt{\left(  1-6\xi\right)  \alpha^{2}+2\xi\alpha\gamma+\alpha
k^{2}-\omega^{2}}}{2\alpha}}\left[  \frac{\Gamma\left(  c_{1}\right)
\Gamma\left(  c_{1}-a_{1}-b_{1}\right)  }{\Gamma\left(  c_{1}-a_{1}\right)
\Gamma\left(  c_{1}-b_{1}\right)  }F\left(  a_{1},b_{1},a_{1}+b_{1}%
+1-c_{1},1-x\right)  \right. \nonumber\\
&  \left.  +\left(  1-x\right)  ^{c_{1}-a_{1}-b_{1}}\frac{\Gamma\left(
c_{1}\right)  \Gamma\left(  a_{1}+b_{1}-c_{1}\right)  }{\Gamma\left(
a_{1}\right)  \Gamma\left(  b_{1}\right)  }F\left(  c_{1}-a_{1},c_{1}%
-b_{1},1+c_{1}-a_{1}-b_{1},1-x\right)  \right]  \ ,
\end{align}
which near infinity, as a function of the radial coordinate $r$, leads to%
\begin{equation}
H\left(  r\right)  \sim_{r\rightarrow\infty}\frac{A_{bh}}{r^{\eta_{+}}}\left(
1+O\left(  \frac{1}{r}\right)  \right)  +\frac{B_{bh}}{r^{\eta_{-}}}\left(
1+O\left(  \frac{1}{r}\right)  \right)  \label{asympbh}%
\end{equation}
where%
\begin{align}
A_{bh}  &  =\frac{\Gamma\left(  c_{1}\right)  \Gamma\left(  c_{1}-a_{1}%
-b_{1}\right)  }{\Gamma\left(  c_{1}-a_{1}\right)  \Gamma\left(  c_{1}%
-b_{1}\right)  }\ ,\label{Abh}\\
B_{bh}  &  =\frac{\Gamma\left(  c_{1}\right)  \Gamma\left(  a_{1}+b_{1}%
-c_{1}\right)  }{\Gamma\left(  a_{1}\right)  \Gamma\left(  b_{1}\right)  }\ ,
\label{Bbh}%
\end{align}
and%
\begin{equation}
\eta_{\pm}=1\pm\sqrt{(1-6\xi)+\frac{(2\gamma\xi+k^{2})}{\alpha}-\frac
{\omega^{2}}{\alpha^{2}}}\ .
\end{equation}

This implies that different modes will have polynomial asymptotic expansions
at infinity in the radial coordinate, with an exponent that is frequency dependent. This is
in contrast with the asymptotically flat case for which $R\left(  r\right)
\sim e^{\pm i\omega r}$, and with the asymptotically AdS case for which
$\eta_{\pm}=\Delta_{\pm}$ being independent of both the angular momentum $k$
and the frequency $\omega$. Since in general $\omega\in\mathbb{C}
$, in order to understand the possible boundary conditions at infinity, we
will require the action principle to attain an extremum on the family of
solutions that are ingoing at the horizon. The action principle leading to
(\ref{scalar}) reads%
\begin{equation}
I=\int d^{4}x\sqrt{-g}\left(  -\frac{1}{2}\nabla_{\mu}\Phi\nabla^{\mu}%
\Phi-\frac{1}{2}\xi R\Phi^{2}\right)  \ ,\label{action}%
\end{equation}
and its on-shell variation with respect to the scalar field leads to the
boundary term%
\begin{equation}
\delta I=-\int_{M}d^{4}x\sqrt{-g}\nabla_{\mu}\left(  \nabla^{\mu}\Phi
\delta\Phi\right)  =-\int_{\partial M}d^{3}x\sqrt{-\gamma}\hat{n}_{\mu}%
\nabla^{\mu}\Phi\delta\Phi\ ,\label{genboundaryterm}%
\end{equation}
where $\gamma$ is the determinant of the induced metric on the boundary, while
$\hat{n}_{\mu}$ is its unit normal vector. The boundary is the union of the
spatial surfaces at $t=t_{i}$ and $t=t_{f}$, with the surface $r=r_{0}$ with
$r_{0}\rightarrow\infty$. As usual, the contribution of the former vanish
since we impose $\delta\Phi\left(  t_{i},r,y\right)  =\delta\Phi\left(
t_{f},r,y\right)  =0$, while the latter leads to%
\begin{equation}
-\lim_{r_{0}\rightarrow\infty}r^{3}\partial_{r}H\delta H|_{r=r_{0}}%
=\lim_{r_{0}\rightarrow\infty}\left.  \frac{\eta_{+}A_{bh}^{2}}{r^{2\eta
_{+}-2}}+\frac{\eta_{-}B_{bh}^{2}}{r^{2\eta_{-}-2}}+\frac{A_{bh}B_{bh}\left(
\eta_{+}+\eta_{-}\right)  }{r^{\eta_{+}+\eta_{-}-2}}\right\vert _{r=r_{0}%
}\delta C_{1}%
\end{equation}
One can check that $\eta_{+}+\eta_{-}-2$ vanishes, while $\operatorname{Re}%
\left(  2\eta_{-}-2\right)  <0$ and $\operatorname{Re}\left(  2\eta
_{+}-2\right)  >0$ on the whole complex $\omega$-plane, therefore in order to
obtain a genuine extremum of the on-shell action principle on the ingoing
solution at the horizon, we need to impose $B_{bh}=0$. From the view point of
the asymptotic expansion (\ref{asympbh}), this corresponds to a Dirichlet
boundary condition. Considering the expression for $B_{bh}$ in (\ref{Bbh}) we
obtain the following two equations for the spectrum%
\begin{align}
a_{1} &  =\frac{1}{2}-\frac{i\omega}{2\alpha}-\frac{\sqrt{2\xi}}{2}%
+\frac{\sqrt{\left(  1-6\xi\right)  \alpha^{2}+2\xi\alpha\gamma+\alpha
k^{2}-\omega^{2}}}{2\alpha}=-p\ \text{with }p=0,1,2,...\ ,\label{spec1}\\
b_{1} &  =\frac{1}{2}-\frac{i\omega}{2\alpha}+\frac{\sqrt{2\xi}}{2}%
+\frac{\sqrt{\left(  1-6\xi\right)  \alpha^{2}+2\xi\alpha\gamma+\alpha
k^{2}-\omega^{2}}}{2\alpha}=-q\ \text{with }q=0,1,2,...\ .\label{spec2}%
\end{align}

Equation (\ref{spec1}) leads to the following purely imaginary spectrum%
\begin{equation}
\omega_{p}=-\frac{\left(  \left(  2\gamma-8\alpha\right)  \xi+2\sqrt{2\xi
}\left(  1+2p\right)  \alpha-4p\left(  1+p\right)  \alpha+k^{2}\right)
\left(  1+2p+2^{1/2}\xi^{1/2}\right)  }{4\xi-2(1+2p)^{2}}i\ ,\label{wpbh}%
\end{equation}
which is a valid solution of (\ref{spec1}) provided%
\begin{equation}
\nu_{p}:=\frac{\left(  1-4\xi+\left(  2p+1\right)  ^{2}\right)  \alpha
+2\gamma\xi+k^{2}-2\left(  2p+1\right)  \alpha\sqrt{2\xi}}{4\left(
1+2p-\sqrt{2\xi}\right)  \alpha}<0\ .\label{neg1}%
\end{equation}
On the other hand, equation (\ref{spec2}) leads to the following set of
frequencies%
\begin{equation}
\omega_{q}=-\frac{((2\gamma-8\alpha)\xi-2\sqrt{2\xi}\left(  1+2q\right)
\alpha-4q(1+q)\alpha+k^{2})(1+2q-\sqrt{2\xi})}{4\xi-2(1+2q)^{2}}%
i\ ,\label{wqbh}%
\end{equation}
which is instead a valid solution of (\ref{spec2}) provided%
\begin{equation}
\nu_{q}:=\frac{\left(  1-4\xi+\left(  2p+1\right)  ^{2}\right)  \alpha
+2\gamma\xi+k^{2}+2\left(  2p+1\right)  \alpha\sqrt{2\xi}}{4\left(
1+2p+\sqrt{2\xi}\right)  \alpha}<0\ .\label{neg2}%
\end{equation}

It can be checked that both spectra (\ref{wpbh}) and (\ref{wqbh}), in the
\textquotedblleft s-wave" case ($k^{2}=0$) lead purely imaginary frequencies
with negative imaginary part. The conditions (\ref{neg1}) and (\ref{neg2})
restrict the values of the non-minimal coupling parameter that lead to non-trivial spectra. An exhaustive exploration of these spectra is beyond the
scope of this work, nevertheless, for the spherically symmetric black holes,
with $k^{2}=l\left(  l+1\right)  =2$ we find a range of values of the
non-minimal coupling $\xi$ leading to unstable modes coming from the spectrum
(\ref{wpbh}) when $p=0$. Figure 1 depicts both $\operatorname{Im}\left(
\omega_{p}\right)  $ and $\nu_{p}$ from (\ref{wpbh}) and (\ref{neg1}),
respectively for a certain range of the non-minimal coupling, showing the
presence of valid modes with $\operatorname{Im}\left(  \omega_{p}\right)  >0$,
therefore unstable. Notice that there is a valid mode for which $\omega=0$,
which can be interpreted as a static scalar cloud \cite{Hod:2012px}. The existence of
these static solutions are usually interpreted as smoking guns for the
existence of a new branch of solutions in which the probe becomes fully
backreacting (see e.g. \cite{Herdeiro:2014goa}). Notice that in our case, the would-be
static backreacting solution might be non-spherically symmetric since $l=1$.
\begin{figure}
\centering
\includegraphics[scale=0.45]{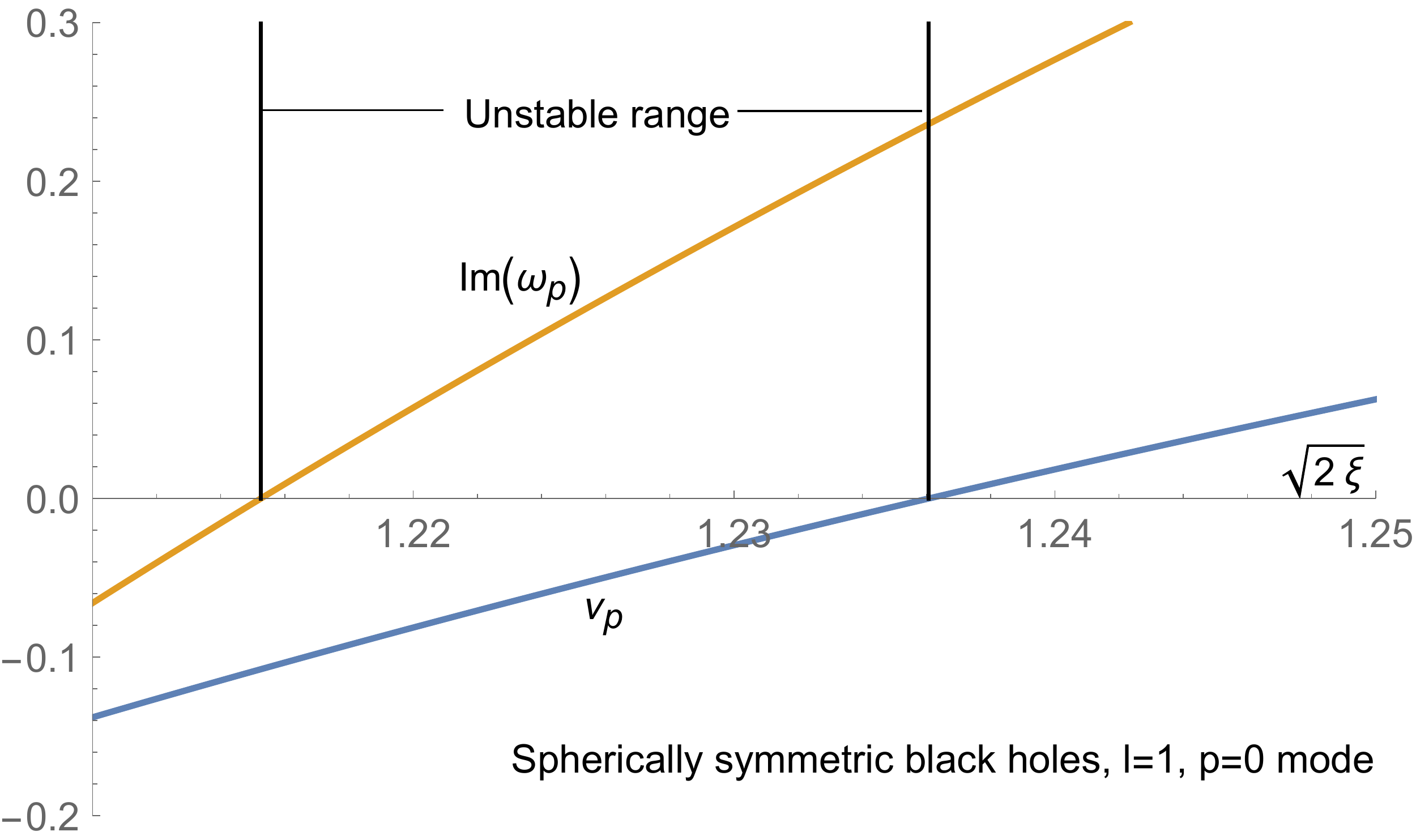}
 \caption{Frequency $\omega_{p}$ and $\nu_{p}$ on the spherically
symmetric black hole $\gamma=1$, for the non-minimally coupled scalar with
non-minimal coupling $\xi$, on the mode with $p=0$ and angular momentum $l=1$.
The allowed modes correspond to values of $\xi$ such that $\nu_{p}<0$. The
modes with frequencies with positive imaginary part are unstable. In
consequence, stability sets an upper bound on the value of the non-minimal
coupling parameter.}
    \label{bhmodes}
\end{figure}

We can see from equations (\ref{neg1}) and (\ref{neg2}) that for a massless,
minimally coupled scalar, namely for $\xi=0$, it is not possible to fulfill
the boundary conditions and there are no quasinormal modes of such massless
scalar probe fields on the black hole background. This situation is similar to
what occurs for a massless scalar probe on the asymptotically locally flat,
static black holes in New Massive Gravity \cite{Anabalon:2019zae}. In the present case, a
non-vanishing value of the non-minimal coupling allows for non-trivial
quasinormal modes, provided (\ref{neg1}) and (\ref{neg2}) are fulfilled. It is
very interesting to notice that such quasinormal frequencies do not depend on
the black hole mass $M=M(r_{+})$, and therefore all the black holes in the family
(\ref{bh}) for different values of $r_{+}$ are isospectral in what regards the
quasinormal modes of the non-minimally coupled scalars. Notice that this is
the case both, in the spherically symmetric and planar cases recovered by
setting $\gamma=1$ and $\gamma=0$, respectively.

\bigskip

It is also illuminating to rewrite the second order equation for the radial
profile of the non-minimally coupled scalar probe in a Schroedinger-like form.
This is achieved by introducing the tortoise coordinate $r_{\ast}$ for the
metric (\ref{bh})%
\begin{equation}
r_{\ast}=\frac{1}{2\alpha}\ln\left(  r^{2}-r_{+}^{2}\right)  \rightarrow
r=\sqrt{r_{+}^{2}+e^{2\alpha r_{\ast}}}\ ,
\end{equation}
which maps $r\in]r_{+},\infty\lbrack$ to the whole real line, i.e. $r_{\ast
}\in]-\infty,+\infty\lbrack$. Notice that we have been able to explicitly
solve $r$ in terms of $r_{\ast}$, which is not possible for Schwarzschild
black hole. Using this fact, we can obtain the potential of the
Schroedinger-like equation explicitly in terms of $r_{\ast}$. Introducing
\begin{equation}
F\left(  r\right)  =\frac{H\left(  r\right)  }{r}\ ,
\end{equation}
leads to%
\begin{equation}
-\frac{d^{2}F}{dr_{\ast}^{2}}+U\left(  r_{\ast}\right)  F=\omega^{2}F\ ,
\end{equation}
with%
\begin{equation}
U\left(  r_{\ast}\right)  =\frac{\alpha\left[  r_{+}^{2}\left(  2\left(
1-4\xi\right)  \alpha+2\gamma\xi+k^{2}\right)  e^{2\alpha r_{\ast}}+e^{4\alpha
r_{\ast}}\left(  \left(  1-6\xi\right)  \alpha+2\gamma\xi+k^{2}\right)
\right]  }{\left(  r_{+}^{2}+e^{2\alpha r_{\ast}}\right)  ^{2}}\ .
\end{equation}
Notice that this potential always vanishes in the near horizon region, namely
when $r_{\ast}\rightarrow-\infty$. Even more, when $\xi=0$ as $r_{\ast}%
\rightarrow\infty$ the potential approaches a positive constant and has a
Heviside-like shape, being a monotonically increasing function of $r_{\ast}$.
As mentioned above, for the minimally coupled case it is impossible to find
quasinormal modes, which is consistent with the basic fact that Schroedinger
equation on a Heviside potential cannot have solutions that approach zero at
$x\rightarrow\infty$ and that represent purely ``outgoing" modes travelling
towards the left as $x\rightarrow-\infty$.

\bigskip

In what follows we move to the problem of computing the spectrum for a
non-minimally coupled scalar probe on the gravitational solitons recently
constructed in \cite{Canfora:2021nca} in $\mathcal{N}=4$ $SU\left(  2\right)  \times SU\left(
2\right)  $ gauged supergravity, both in the supersymmetric and
non-supersymmetric cases.

\bigskip

\section{Spectrum of probe scalars on solitons}

As shown in \cite{Canfora:2021nca}, $\mathcal{N}=4$\ $SU(2)\times SU\left(  2\right)  $ gauged
supergravity has the following soliton solution%
\begin{equation}
ds^{2}=-\rho dt^{2}+g\left(  \rho\right)  d\varphi^{2}+\frac{d\rho^{2}%
}{g\left(  \rho\right)  }+\rho dy^{2}\ ,\label{soliton}%
\end{equation}
where%
\begin{equation}
g\left(  \rho\right)  =\alpha\left(  \rho-m-\frac{q^{2}}{\rho}\right)  \ ,
\end{equation}
$m$ and $q$ being integration constants, $\alpha$ is related with the gauge
couplings and $\varphi$ is identified with period $\beta_{\varphi}$ given by%
\begin{equation}
\beta_{\varphi}=\frac{4\pi}{g^{\prime}\left(  \rho_{0}\right)  }\ .
\end{equation}
Here $g\left(  \rho_{0}\right)  =0$, $\rho\geq\rho_{0}$ and the constants
$\alpha$, $q$, the gauge fields and the dilaton are given by%
\begin{align}
\alpha &  =\frac{1}{2}\left(  e_{A}^{2}+e_{B}^{2}\right)  \ ,\qquad
q^{2}=\frac{8\left(  Q_{A}^{2}+Q_{B}^{2}\right)  }{e_{A}^{2}+e_{B}^{2}}\ ,\\
A_{\left[  1\right]  }^{i} &  =\frac{Q_{A}}{\rho}d\varphi\delta_{3}%
^{i}\ ,\qquad \,\quad B_{\left[  1\right]  }^{i}=\frac{Q_{B}}{\rho}d\varphi\delta
_{3}^{i}\ ,\\
\phi\left(  r\right)   &  =-\frac{1}{2}\ln\rho\ .
\end{align}
For general values of the integration constants $m$ and $q$, the non-minimally
coupled scalar probe does not admit a solution in a closed form. Nevertheless,
for the case $q=0$ and $m$ arbitrary, as well as for the case $m=0$ and $q$
arbitrary, the non-minimally coupled scalar field can indeed be solved in
terms of hypergeometric functions, consequently boundary conditions can be
imposed in a closed manner, leading to a discrete set of frequencies.
Hereafter we refer to these special cases as soliton-1 and soliton-2, which
are defined by the metric (\ref{soliton}), with $g\left(  \rho\right)  $ given
by%
\begin{align}
g_{sol1}\left(  \rho\right)   &  =\alpha\left(  \rho-m\right)  \ ,\label{sol1}%
\\
g_{sol2}\left(  \rho\right)   &  =\alpha\left(  \rho-\frac{q^{2}}{\rho
}\right)  \ ,\label{sol2}%
\end{align}
respectively.The soliton-2 spacetime leads to
a supersymmetric configuration that preserves 1/4 of the supersymmetry \cite{Canfora:2021nca}.

\bigskip

Defining $\varphi=\frac{\beta_{\varphi}}{2\pi}\phi$, the coordinate $\phi$
will have period $2\pi$, and the metric (\ref{soliton}) reduces to%
\begin{equation}
ds^{2}=-\rho dt^{2}+\frac{\beta_{\varphi}^{2}}{4\pi^{2}}g\left(  \rho\right)
d\phi^{2}+\frac{d\rho^{2}}{g\left(  \rho\right)  }+\rho dy^{2}\ .\label{sols}%
\end{equation}
Given the isometries of this spacetime we write the following separation
ansatz for a scalar probe%
\begin{equation}
\Phi=\operatorname{Re}\left(  \sum_{n}\int d\omega dke^{-i\omega t+iky+in\phi
}H_{\omega,k,n}\left(  \rho\right)  \right)  \ .
\end{equation}
The Ricci scalar of (\ref{sols}) has a non-trivial profile and it is given by%
\begin{equation}
R=\frac{g\left(  \rho\right)  }{2\rho^{2}}-g^{\prime\prime}\left(
\rho\right)  -\frac{2g^{\prime}\left(  \rho\right)  }{\rho}\ .
\end{equation}
Introducing the notation $H_{\omega,k,n}\left(  \rho\right)  =H\left(
\rho\right)  $, the equation for the non-minimally coupled scalar%
\begin{equation}
\square\Phi-\xi R\Phi=0\ ,
\end{equation}
leads to the following ODE for the radial dependence%
\begin{equation}
2\rho^{2}g^{2}\beta_{\varphi}^{2}H^{\prime\prime}+2g\rho\beta_{\varphi}%
^{2}\left(  g\rho\right)  ^{\prime}H^{\prime}+\left(  g\beta_{\varphi}%
^{2}\left(  2g^{\prime\prime}\rho^{2}+4g^{\prime}\rho-g\right)  \xi
-2\rho\left(  4\pi^{2}n^{2}\rho+g\beta_{\varphi}^{2}\left(  k^{2}-\omega
^{2}\right)  \right)  \right)  H=0\ .
\end{equation}
Here the prime denotes derivative with respect to $\rho$. In what follows we
analyze this equation for both soliton-1 and soliton-2 spacetimes, separately.

\subsection{Non-supersymmetric soliton}

For the family of solitons defined by the function soliton-1 in (\ref{sol1}),
we have $\rho_{0}=m$, and\ $g^{\prime}\left(  \rho_{0}\right)  =\alpha$,
therefore $\beta_{\varphi}=\frac{4\pi}{\alpha}$. Introducing the coordinate
$x$ such that%
\begin{equation}
\rho=\frac{\rho_{0}}{1-x}\ ,
\end{equation}
which maps $\rho\in\lbrack\rho_{0},\infty\lbrack$ to $x\in\lbrack0,1[$, leads
to an equation for the radial profile that can be integrated in terms of
hypergeometric functions. Imposing regularity at the origin $\rho=\rho_{0}$
($x=0$) leads to the following solution%
\begin{equation}
H\left(  \rho\left(  x\right)  \right)  =C_{1}x^{\frac{|n|}{2}}\left(
1-x\right)  ^{\frac{1}{2}\left(  1-\sqrt{\left(  1-6\xi\right)  +n^{2}%
+\frac{4\left(  k^{2}-\omega^{2}\right)  }{\alpha}}\right)  }F\left(
\alpha_{1},\beta_{1},\gamma_{1},x\right)  \ ,\label{Hsol1}%
\end{equation}
with%
\begin{align}
\alpha_{1} &  =\frac{1}{2}\left(  1+|n|+\sqrt{2\xi}\right)  -\frac{1}{2}%
\sqrt{(1-6\xi+n^{2})+\frac{4\left(  k^{2}-\omega^{2}\right)  }{\alpha}}\ ,\\
\beta_{1} &  =\frac{1}{2}\left(  1+|n|-\sqrt{2\xi}\right)  -\frac{1}{2}%
\sqrt{(1-6\xi+n^{2})+\frac{4\left(  k^{2}-\omega^{2}\right)  }{\alpha}}\ ,\\
\gamma_{1} &  =1+|n|\ .\nonumber
\end{align}
As in the previous section, using Kummer identities allows to rewrite
(\ref{Hsol1}) as%
\begin{align}
H\left(  \rho\left(  x\right)  \right)   &  =C_{1}x^{\frac{|n|}{2}}\left(
1-x\right)  ^{\frac{1}{2}\left(  1-\sqrt{\left(  1-6\xi\right)  +n^{2}%
+\frac{4\left(  k^{2}-\omega^{2}\right)  }{\alpha}}\right)  }\times\nonumber\\
&  \left[  \frac{\Gamma\left(  \gamma_{1}\right)  \Gamma\left(  \gamma
_{1}-\alpha_{1}-\beta_{1}\right)  }{\Gamma\left(  \gamma_{1}-\alpha
_{1}\right)  \Gamma\left(  \gamma_{1}-\beta_{1}\right)  }F\left(  \alpha
_{1},\beta_{1},\alpha_{1}+\beta_{1}+1-\gamma_{1},1-x\right)  \right.  \\
&  \left.  +\left(  1-x\right)  ^{\gamma_{1}-\alpha_{1}-\beta_{1}}\frac
{\Gamma\left(  \gamma_{1}\right)  \Gamma\left(  \alpha_{1}+\beta_{1}%
-\gamma_{1}\right)  }{\Gamma\left(  \alpha_{1}\right)  \Gamma\left(  \beta
_{1}\right)  }F\left(  \gamma_{1}-\alpha_{1},\gamma_{1}-\beta_{1},1+\gamma
_{1}-\alpha_{1}-\beta_{1},1-x\right)  \right]
\end{align}
which leads to the following two leading terms on each branch of the
asymptotic behavior as $x\rightarrow1$%
\begin{equation}
H\left(  x\right)  \underset{x\rightarrow1}{\sim}A_{1}\left(  1-x\right)
^{\delta_{-}}+B_{1}\left(  1-x\right)  ^{\delta+}\ ,\label{asympsol1}%
\end{equation}
with%
\begin{equation}
\delta_{\pm}=\frac{1}{2}\left(  1\pm\sqrt{\left(  1-6\xi\right)  +n^{2}%
+\frac{4\left(  k^{2}-\omega^{2}\right)  }{\alpha}}\right)  \ ,
\end{equation}
and%
\begin{align}
A_{1} &  =\frac{\Gamma\left(  \gamma_{1}\right)  \Gamma\left(  \gamma
_{1}-\alpha_{1}-\beta_{1}\right)  }{\Gamma\left(  \gamma_{1}-\alpha
_{1}\right)  \Gamma\left(  \gamma_{1}-\beta_{1}\right)  }\ ,\\
B_{1} &  =\frac{\Gamma\left(  \gamma_{1}\right)  \Gamma\left(  \alpha
_{1}+\beta_{1}-\gamma_{1}\right)  }{\Gamma\left(  \alpha_{1}\right)
\Gamma\left(  \beta_{1}\right)  }\ .
\end{align}
Since the exponents in the asymptotic behavior (\ref{asympsol1}) are $\omega$
dependent, we must be careful when imposing the boundary conditions. Again,
the boundary term coming from the on-shell variation of the action principle
(\ref{action})-(\ref{genboundaryterm}) leads to a single contribution at
infinity coming from the surface $x=x_{0}\rightarrow1$. In terms of the
coordinate $x$, the non-supersymmetric soliton spacetime reads%
\begin{equation}
ds^{2}=-\frac{\rho_{0}}{1-x}dt^{2}+\frac{4\rho_{0}x}{1-x}d\phi^{2}+\frac
{\rho_{0}}{\alpha x\left(  1-x\right)  ^{3}}dx^{2}+\frac{\rho_{0}}{1-x}%
dy^{2}\ ,
\end{equation}
while the boundary term of the on-shell variation of the action reads%
\begin{align}
&  \lim_{r\rightarrow\infty}\int d^{3}x\sqrt{-\gamma}\hat{n}_{\mu}\nabla^{\mu
}\Phi\delta\Phi\nonumber\\
&  \sim\lim_{x\rightarrow1}\left(  \delta_{-}^{2}A_{1}^{2}\left(  1-x\right)
^{2\delta_{-}-1}+\delta_{+}^{2}B_{1}^{2}\left(  1-x\right)  ^{2\delta_{+}%
-1}+A_{1}B_{1}\left(  \delta_{+}+\delta_{-}\right)  \left(  1-x\right)
^{\delta_{+}+\delta_{-}-1}\right)  \delta C_{1}%
\end{align}

It can be checked that $\operatorname{Re}\left(  2\delta_{-}-1\right)  <0$ on
the whole complex $\omega$-plane, while $\operatorname{Re}\left(  2\delta
_{+}-1\right)  >0$ and $\delta_{+}+\delta_{-}-1=0$. Therefore, in order to
make the boundary term to vanish when evaluated on-shell on the branch that is
regular at the origin, we must impose%
\begin{equation}
A_{1}=\frac{\Gamma\left(  \gamma_{1}\right)  \Gamma\left(  \gamma_{1}%
-\alpha_{1}-\beta_{1}\right)  }{\Gamma\left(  \gamma_{1}-\alpha_{1}\right)
\Gamma\left(  \gamma_{1}-\beta_{1}\right)  }=0\ .
\end{equation}
Notice that this is actually a Dirichlet boundary condition as can be seen
from (\ref{asympsol1}). The spectrum is therefore obtained from%
\begin{align}
\gamma_{1}-\alpha_{1} &  =\frac{\sqrt{(1-6\xi+n^{2})\alpha+4\left(
k^{2}-\omega^{2}\right)  }+\sqrt{\alpha}\left(  1+|n|-\sqrt{2\xi}\right)
}{2\sqrt{\alpha}}=-p\ ,\label{specsol11}\\
\gamma_{1}-\beta_{1} &  =\frac{\sqrt{(1-6\xi+n^{2})\alpha+4\left(
k^{2}-\omega^{2}\right)  }+\sqrt{\alpha}\left(  1+|n|+\sqrt{2\xi}\right)
}{2\sqrt{\alpha}}=-q\ ,\label{specsol12}%
\end{align}
with $q$ and $p$ in $\left\{  0,1,2,...\right\}  $. One can also check that
the second quantization condition (\ref{specsol12}) cannot be fulfilled,
nevertheless the quantization condition (\ref{specsol11}) leads to the
spectrum%
\begin{equation}
\omega_{p}=\pm\frac{1}{2}\sqrt{2\alpha\left(  |n|+2p+1\right)  \sqrt{2\xi
}+4k^{2}-2\left(  |n|\left(  1+2p\right)  +4\xi+2p\left(  1+p\right)  \right)
\alpha}\ ,\label{wsol1}%
\end{equation}
which is a valid solution of (\ref{specsol11}) provided%
\begin{equation}
\nu_{ns}:=|n|+1-\sqrt{2\xi}+2p<0\ .
\end{equation}
As in the case of the black hole, for the non-supersymmetric soliton requiring
regularity at the origin and Dirichlet boundary condition at infinity leads to
an eigenvalue problem with a void spectrum when $\xi=0$. Nevertheless, the presence of the
non-minimal coupling leads to non-trivial probe modes.

The spectrum of the scalar on the non-supersymmetric soliton can be of diverse
nature. Depending on the values of the parameters, it could be void, purely
oscillatory namely with real frequencies (\ref{wsol1}) or unstable. The
different behavior can be seen as separated by thresholds in the value of the
non-minimal coupling $\xi$. Figure 2 shows two possible spectra.

\begin{figure}
\centering
\includegraphics[scale=0.3]{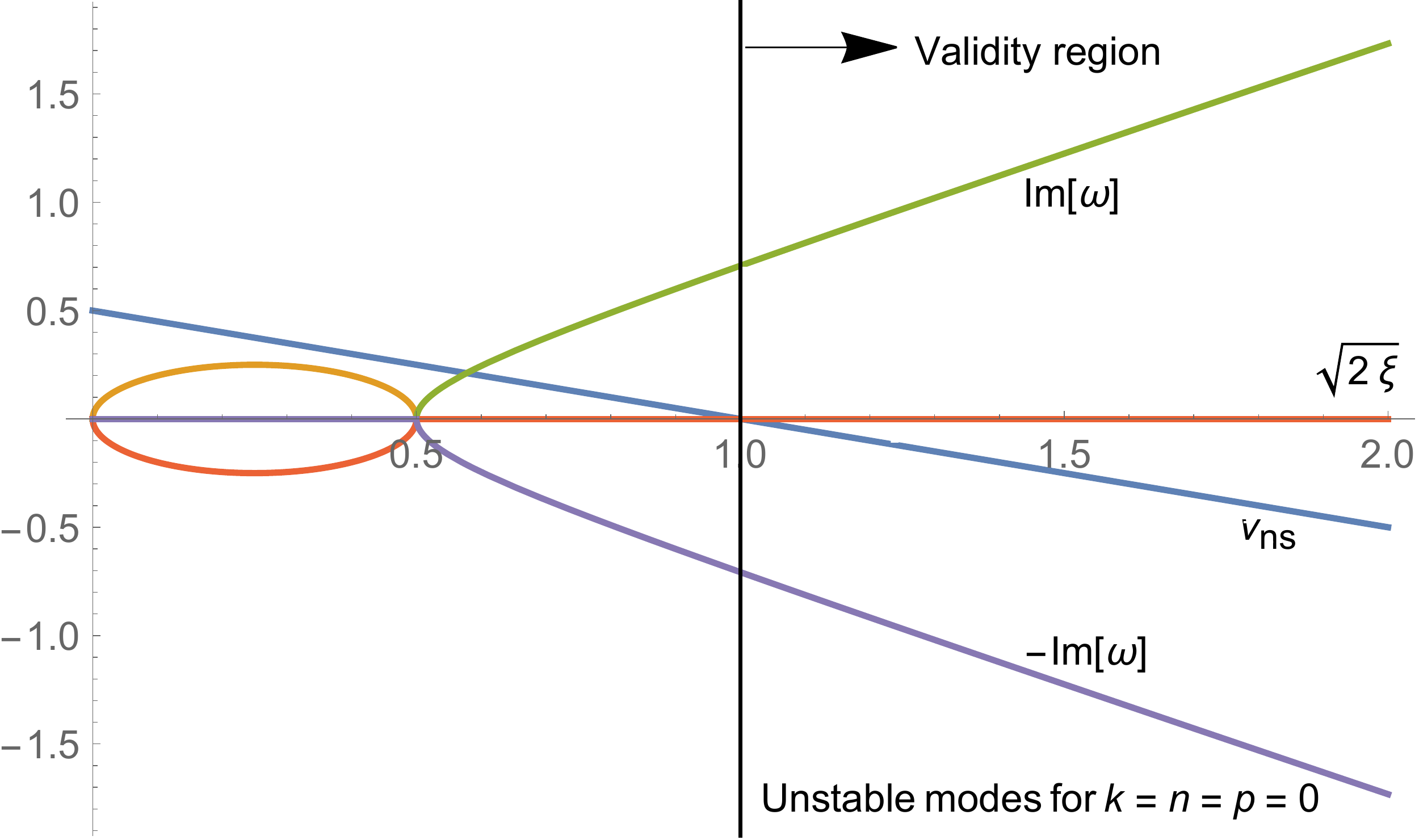}\qquad \includegraphics[scale=0.3]{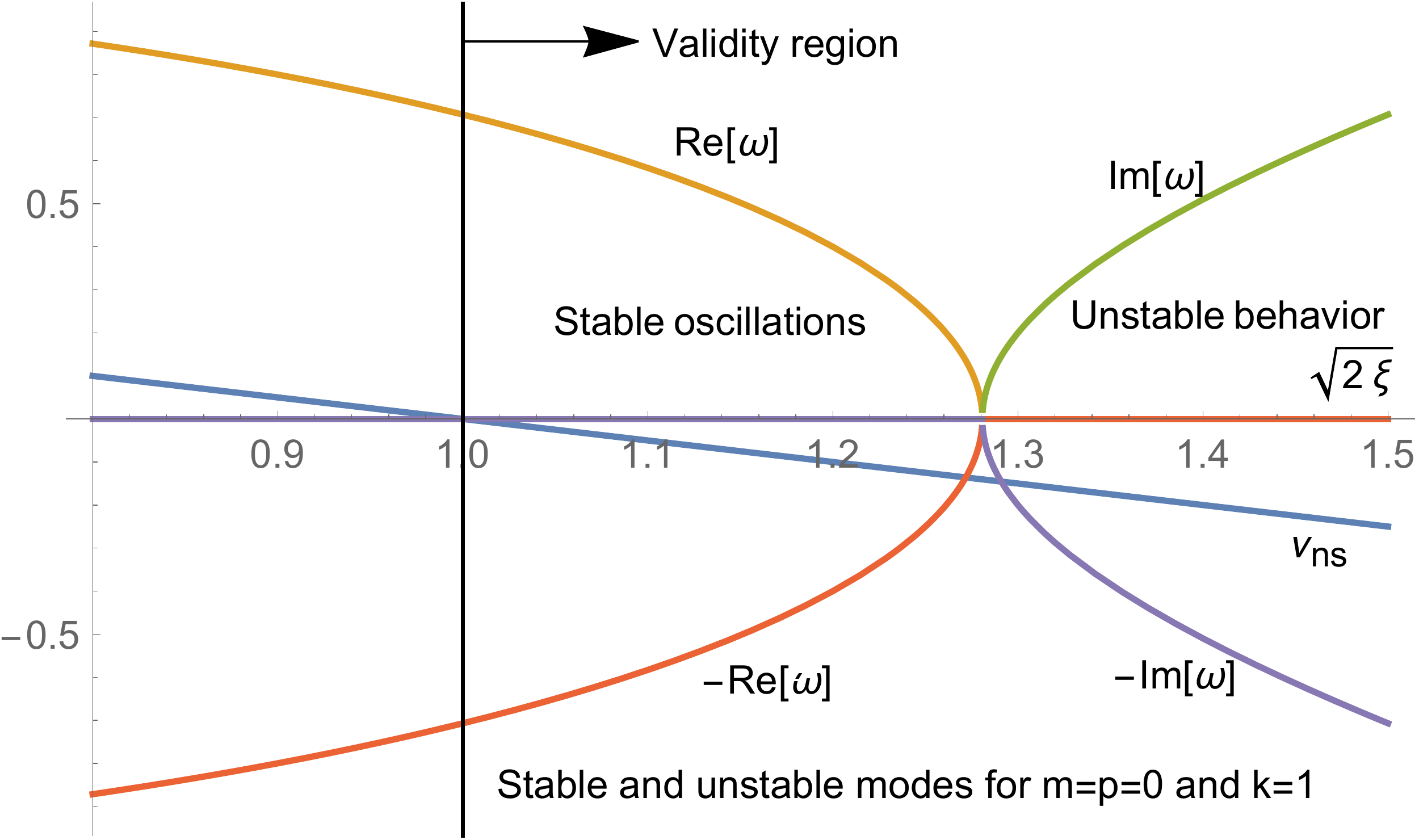}
 \caption{The panels show the spectra of stable and unstable modes of the non-minimally coupled scalar probe on the background of the non-supersymmetric soliton. Valid solutions for the quantization equation leading to the frequencies require $\nu_p<0$, therefore in both panels, to the left of the vertical black line, there are no allowed modes given our boundary conditions.}
    \label{nssplots}
\end{figure}

\subsection{Supersymmetric soliton}

The 1/4 supersymmetric soliton is
given by the metric (\ref{sols}) with the function $g\left(  \rho\right)  $
given by%
\begin{equation}
g_{sol2}\left(  \rho\right)  =\alpha\left(  \rho-\frac{q^{2}}{\rho}\right)
\ .
\end{equation}
In this case the smooth origin of the spacetime is located at $\rho=\rho
_{0}=q$ and the equation for the radial profile of the non-minimally coupled
scalar probe has the following solution which is regular at the origin%
\begin{equation}
H\left(  x\right)  =C_{1}x^{\frac{|n|}{2}}\left(  1-x\right)  ^{\frac{1}%
{4}-\frac{1}{4}\sqrt{\left(  1-6\xi\right)  +4n^{2}+\frac{\left(
k^{2}-4\omega^{2}\right)  }{\alpha}}}F\left(  \alpha_{2},\beta_{2},\gamma
_{2},x\right)  \ , \label{Hsol2}%
\end{equation}
where in this case the coordinate $x$ is conveniently chosen as%
\begin{equation}
\rho=\frac{\rho_{0}}{\left(  1-x\right)  ^{1/2}}\ .
\end{equation}
Here, the parameters of the hypergeometric function in (\ref{Hsol2}) are given
by%
\begin{align}
\alpha_{2}  &  =-\frac{1}{4}\sqrt{1-6\xi+4n^{2}+\frac{4\left(  k^{2}%
-\omega^{2}\right)  }{\alpha}}+\frac{1}{2}\left(  |n|+1+\frac{1}{2}%
\sqrt{1+2\xi}\right)  \ ,\\
\beta_{2}  &  =-\frac{1}{4}\sqrt{1-6\xi+4n^{2}+\frac{4\left(  k^{2}-\omega
^{2}\right)  }{\alpha}}+\frac{1}{2}\left(  |n|+1-\frac{1}{2}\sqrt{1+2\xi
}\right)  \ ,\\
\gamma_{2}  &  =1+|n|\ .
\end{align}

Using Kummer identity in (\ref{Hsol2}) leads to the following leading terms of
the two branches of asymptotic behavior%
\begin{equation}
H\left(  x\right)  \sim_{x\rightarrow1}A_{2}\left(  1-x\right)  ^{\lambda_{-}%
}+B_{2}\left(  1-x\right)  ^{\lambda_{+}}\ ,
\end{equation}
with%
\begin{equation}
\lambda_{\pm}=\frac{1}{4}\pm\frac{1}{4}\sqrt{\left(  1-6\xi\right)
+4n^{2}+\frac{\left(  k^{2}-4\omega^{2}\right)  }{\alpha}}\ ,
\end{equation}
and%
\begin{align}
A_{2} &  =\frac{\Gamma\left(  \gamma_{2}\right)  \Gamma\left(  \gamma
_{2}-\alpha_{2}-\beta_{2}\right)  }{\Gamma\left(  \gamma_{2}-\alpha
_{2}\right)  \Gamma\left(  \gamma_{2}-\beta_{2}\right)  }\ ,\nonumber\\
B_{2} &  =\frac{\Gamma\left(  \gamma_{2}\right)  \Gamma\left(  \alpha
_{2}+\beta_{2}-\gamma_{2}\right)  }{\Gamma\left(  \alpha_{2}\right)
\Gamma\left(  \beta_{2}\right)  }\ .
\end{align}
As in the previous section, when the variation of the action is evaluated on
the solution that is regular at the origin, one obtains a boundary term that
vanishes iff%
\begin{equation}
A_{2}=\frac{\Gamma\left(  \gamma_{2}\right)  \Gamma\left(  \gamma_{2}%
-\alpha_{2}-\beta_{2}\right)  }{\Gamma\left(  \gamma_{2}-\alpha_{2}\right)
\Gamma\left(  \gamma_{2}-\beta_{2}\right)  }=0\ .
\end{equation}
In consequence, this implies the following two quantization conditions for the
spectrum%
\begin{align}
\gamma_{2}-\alpha_{2} &  =\frac{1}{2}+\frac{|n|}{2}+\frac{1}{4}\sqrt
{1-6\xi+4n^{2}+\frac{4\left(  k^{2}-\omega^{2}\right)  }{\alpha}}-\frac{1}%
{4}\sqrt{1+2\xi}=-p\ ,\label{spec21}\\
\gamma_{2}-\beta_{2} &  =\frac{1}{2}+\frac{|n|}{2}+\frac{1}{4}\sqrt
{1-6\xi+4n^{2}+\frac{4\left(  k^{2}-\omega^{2}\right)  }{\alpha}}+\frac{1}%
{4}\sqrt{1+2\xi}=-q\ ,
\end{align}
with $p$ and $q$ elements of $\left\{  0,1,2,3,...\right\}  $. It can be shown
that the second condition cannot be fulfilled, while the former leads to the
spectrum%
\begin{equation}
\omega=\pm\sqrt{k^{2}+\left(  1+|n|+2p\right)  \alpha\sqrt{1+2\xi}%
-\alpha\left(  2\xi+\left(  1+2p\right)  ^{2}+2|n|\left(  1+2p\right)
\right)  }\ ,
\end{equation}
which are genuine solutions of (\ref{spec21}) provided%
\begin{equation}
\nu_{susy}=2+|n|+4p-\sqrt{1+2\xi}<0
\end{equation}

Depending on the ranges of the parameters one can obtain the same qualitative
spectra as in the non-supersymmetric soliton, namely there is a range of values for the non-minimal coupling for
which the spectrum is void, while in the complementary range one can have both
stable and unstable modes. Stable oscillatory behavior can be achieved
provided one restricts the values of the non-minimal coupling.

\section{Final remarks}
In this paper we have found that $\mathcal{N}=4$ $SU\left(  2\right)  \times
SU\left(  2\right)  $ gauged supergravity admits black holes an solitons with sufficiently simple geometry that allows to compute the spectrum of a
non-minimal scalar probe in an exact manner. The spacetimes approach a
background at infinity which is not maximally symmetric, but possesses and
extra conformal Killing vector, which is due to the fact that the dilatonic
potential of the theory does not have local extrema. As reported in \cite{Canfora:2021nca} the
solitonic geometries are smooth at the origin and can preserve $1/4$ of the
supersymmetries and can be obtained from the corresponding planar black holes
via a double analytic continuation, as it is the case of the recently reported
solitons in $\mathcal{N}=2$ gauged supergravity in four dimensions \cite{Anabalon:2021tua}. At
the origin and in the near horizon region, the boundary conditions are clear
and are given by regularity and purely ingoing modes, respectively. Due to
the non-trivial geometry at infinity, the behavior of the scalar probe in the
asymptotic region is given by powers of the radial coordinate which depend on
the frequencies. In order to select a consistent boundary condition at
infinity we impose that the on-shell variation of the action functionals must
vanish. This leads to a Dirichlet boundary condition and allows to write the
spectra in a closed form. For the massless scalar probe it is impossible to
fulfill these boundary conditions. For the black
holes, this is consistent with the fact that the effective Schroedinger-like
potential controlling the radial dependence of the scalar probe in terms of
the tortoise coordinate, has a Heaviside function shape. Including a
non-minimal coupling allows for a non-trivial spectrum which surprisingly, in
the case of the black hole, does not depend on the value of the mass of the
spacetime. Therefore all these geometries are isospectral in what regards to
the non-minimally coupled wave operator. Given the integrability properties of
this potential it will be interesting to compare our results with the recently
reported potentials coming from a geometric approach to spectral theory in
connection with $SU\left(  2\right)  $ Seiberg-Witten theory with fundamental
hypermultiplets (see Section 2 of \cite{Aminov:2020yma}). Such potential  are also given in terms of ratios of linear combinations of exponentials.

Stability of the modes is achieved for a certain range of non-minimal
couplings, above which one finds modes that are exponentially growing in time,
and that are in consequence unstable. The stable and unstable regimes are
separated by solutions to the boundary eigenvalue problem which are time
independent. These solutions have the same properties as the scalar clouds
found in \cite{Hod:2012px} which from the point of view of the fully backreacting theory
are branching spacetimes to a new family of solutions (see e.g. \cite{Herdeiro:2014goa}).

We have been able to solve in a closed form the non-minimally coupled scalar
probe on a family of black holes and solitons of the Freedman-Schwarz model,
even in the case of 1/4-BPS geometries. Such scalar probe goes beyond the
field content of the theory, and it would be interesting to see whether some
of the exact results we have obtained here, are also present in the context of
gravitational perturbation theory considering only the fields that lead to the
supersymetric model even if one has to rely on numerical or perturbative methods. We expect to report along these lines in the near future.

\section*{Acknowledgements}

We thank Andr\'{e}s Anabal\'{o}n and Fabrizio Canfora for useful comments.
This work is partially funded by Beca ANID de Mag\'{\i}ster 22201618 and
FONDECYT grants 1181047. J.O. also thanks the support of Proyecto de
Cooperaci\'{o}n Internacional 2019/13231-7 FAPESP/ANID.

\end{document}